\documentstyle[seceq,epsfig]{ptptex}

\title{Tensor Product Variational Formulation for Quantum Systems}

\author
{ Yukinobu {\sc Nishio}, Nobuya {\sc Maeshima}$^{1}$, 
Andrej {\sc Gendiar}$^{2}$, and 
Tomotoshi {\sc Nishino}\footnote{E-mail address:
nishino@phys.sci.kobe-u.ac.jp} }

\inst
{Department of Physics, Graduate School of Science,
Kobe University, rokkodai 657-8501\\
$^{1}$ Cybermedia Center,
Osaka University, Machikaneyama 1-43, Osaka 560-0043\\
$^{2}$ Institute of Electrical Engineering, Slovak Academy of Sciences,
D\'ubravsk\'a cesta 9, SK-842 39 Bratislava, Slovakia}

\recdate{}

\abst{
We consider a variational problem for the two-dimensional (2D) Heisenberg 
and XY models, using a trial state which is constructed as a 2D product of local 
weights. Variational energy is calculated by use of the the corner transfer matrix renormalization group (CTMRG) method, and its upper bound is surveyed. 
The variational approach is a way of applying
the density matrix renormalization group method (DMRG) to 
infinite size 2D quantum systems. 
}

\begin{document}

\maketitle

\section{Introduction}
As a precise numerical method for one-dimensional (1D) quantum systems,
the density matrix renormalization group (DMRG) has been widely applied 
to unsolved problems in condensed matter physics.~\cite{Wh1,Wh2,rev,Shibata} 
If the linear dimension of the system is not so large ($\sim$ of the order of ten),
the method is also applicable to 2D quantum models by use of a
mapping from a finite size 2D cluster to a 1D chain that contains long-range 
interactions.~\cite{Liang,Xiang} 

On the other hand, it is difficult to apply DMRG 
to infinitely large 2D quantum systems. Apparently the mapping from 2D lattice 
to 1D chain is inapplicable if the system size is infinite. In addition, the decay of the 
density matrix eigenvalue is very slow in higher dimensions,~\cite{Peshel1,Peshel2}  
and the phenomenon prevents to obtain a good renormalization group transformation 
that maps half (or quarter) infinite 2D system into a block spin. \footnote{Existence 
of a good renormalization group transformation is not excluded, in case that 
the transformation is
created some method other than the diagonalization of the density matrix.} A naive 
extension of DMRG formulation to higher dimension shall encounter very poor numerical
results, as was reported in an application of DMRG to 3D classical systems.~\cite{CTTRG}

Stepping back to the DMRG formulation for 1D systems, the numerical efficiency 
of DMRG 
partially comes from its variational structure, where a trial state is represented 
as a product of orthogonal matrices.~\cite{Ost,Rom,Sierra,And} It is 
possible to extend such a construction of variational state to 2D 
quantum and 3D classical systems. For these higher dimensional systems one
has to prepare a trial states that are represented as 2D product of local weights. 
For example, Mart{\'\i}n-Delgad, {\it et al.} employed the 6-vertex model as a trial state 
for 2D lattice spin/electron systems.~\cite{Sierra2} Okunishi and Nishino
considered an extension of the Kramers-Wannier approximation~\cite{KWA}
to the 3D Ising model, representing the trial state as the 2D Ising model
under magnetic field,~\cite{KWON} or more generally as the interaction
round a face (IRF) model.~\cite{KWON2} Let us call these variational approaches
as `the tensor product variational approach' (TPVA) in the following.

In this paper we investigate the numerical efficiency of TPVA when it is 
applied to the square lattice $S = 1/2$ Heisenberg and XY models. 
We employ an isotropic and uniform IRF model as the trial state, that contains 3 adjustable parameters.  Our trial state includes 
the one parameter variational state proposed by Suzuki and Miyashita for 
the study of square lattice XY model.~\cite{Suz,Miy} The local construction
of the variational state enables us to numerically calculate the energy expectation value
by use of DMRG~\cite{Hieida} or by Monte Carlo simulations.~\cite{Zitt2}

In the next section we explain the construction of the trial state, and the way of
calculating the energy expectation value. We show the
numerical result in \S 3, and discuss a way of improving the variational formulation
in the last section.

\section{Construction of the Variational State}

We consider the $S = 1/2$ XXZ model on the square lattice as an
example of 2D quantum systems. Its Hamiltonian is represented as
\begin{equation}
H = \sum_{\{ {\bf r} {\bf r}' \}}^{~} \left( 
S_{{\bf r}}^{\rm x} S_{{\bf r}'}^{\rm x} +
S_{{\bf r}}^{\rm y} S_{{\bf r}'}^{\rm y} + \alpha
S_{{\bf r}}^{\rm z} S_{{\bf r}'}^{\rm z}
\right) \, ,
\end{equation}
where $\{ {\bf r} {\bf r}' \}$ denotes pairs of neighboring lattice sites. The 
parameter $\alpha$ is chosen to be either 1 (the Heisenberg model) or zero (the XY model). 
Since the lattice is bipartite, the following Hamiltonian
\begin{equation}
{\tilde H} = \sum_{\{ {\bf r} {\bf r}' \}}^{~} \left( -
S_{{\bf r}}^{\rm x} S_{{\bf r}'}^{\rm x} -
S_{{\bf r}}^{\rm y} S_{{\bf r}'}^{\rm y} + \alpha
S_{{\bf r}}^{\rm z} S_{{\bf r}'}^{\rm z}
\right) \, ,
\end{equation}
has the same same energy spectrum as $H$. In the following we treat ${\tilde H}$ 
instead of $H$ for the purpose of simplifying the variational 
formulation.\footnote{One has to prepare two different local weights for $H$ and multiply
them alternatively when constructing the variational state. Note
that the mapping from $H$ to ${\tilde H}$ has nothing to do with the elimination of negative
sign from the local weight.}
Our interest is in finding out a good variational function that minimizes the
energy expectation value
\begin{equation}
\Lambda = \frac{\langle \Psi | {\tilde H} | \Psi \rangle}{\langle \Psi | \Psi \rangle} 
\end{equation}
within the restriction where $| \Psi \rangle$ is represented as a product of local weights.

Let us introduce a notation $\sigma_{\bf r}^{~} = 2 S_{\bf r}^z = \pm 1$, and
write the trial wave function as 
 $\Psi( \{ \sigma \} ) \equiv \langle \{ \sigma \} | \Psi \rangle$, 
where $\{ \sigma \}$ represents a spin configuration of all the spins on the 2D lattice. 
We employ a trial wave function in the form of the uniform product of local weights
\begin{equation}
\Psi( \{ \sigma \} ) = \prod_{\bf r}^{~} \,\, W\left(
\begin{array}{cc} 
\sigma_{{\bf r}+{\hat {\bf j}}} & \sigma_{{\bf r}+{\hat {\bf i}}+{\hat {\bf j}}}\\ 
\sigma_{{\bf r}~~~}^{~} & \sigma_{{\bf r}+{\hat {\bf i}}~~~}
\end{array}
\right) \, .
\end{equation}
The weight $W$ is a function of 4 neighboring spin variables
$\sigma_{{\bf r}}^{~}$, $\sigma_{{\bf r}+{\hat {\bf i}}}$, 
$\sigma_{{\bf r}+{\hat {\bf j}}}$, and 
$\sigma_{{\bf r}+{\hat {\bf i}}+{\hat {\bf j}}}$ where ${\hat {\bf i}}$ and 
${\hat {\bf j}}$ represent the unit lattice vector to X and Y directions, respectively.
According to the symmetries of the Hamiltonian ${\tilde H}$, the local weight has
only 3 independent parameters, that are
\begin{eqnarray}
1 &=& 
W\left( \begin{array}{cc} + & - \\ - & +  \end{array} \right) =
W\left( \begin{array}{cc} - & + \\ + & -  \end{array} \right)\nonumber\\
a &=&
W\left( \begin{array}{cc} + & + \\ - & -  \end{array} \right) =
W\left( \begin{array}{cc} + & - \\ + & -  \end{array} \right) =
W\left( \begin{array}{cc} - & - \\ + & +  \end{array} \right) =
W\left( \begin{array}{cc} - & + \\ - & +  \end{array} \right)\nonumber\\
b &=&
W\left( \begin{array}{cc} + & - \\ - & -  \end{array} \right) =
W\left( \begin{array}{cc} - & + \\ - & -  \end{array} \right) =
W\left( \begin{array}{cc} - & - \\ - & +  \end{array} \right) =
W\left( \begin{array}{cc} - & - \\ + & -  \end{array} \right)\nonumber\\
&=&
W\left( \begin{array}{cc} - & + \\ + & +  \end{array} \right) =
W\left( \begin{array}{cc} + & - \\ + & +  \end{array} \right) =
W\left( \begin{array}{cc} + & + \\ + & -  \end{array} \right) =
W\left( \begin{array}{cc} + & + \\ - & +  \end{array} \right)\nonumber\\
c &=& 
W\left( \begin{array}{cc} - & - \\ - & -  \end{array} \right) =
W\left( \begin{array}{cc} + & + \\ + & +  \end{array} \right) \, ,
\end{eqnarray} 
where we have written up and down spins by '$+$' and '$-$', respectively.
Since we have constructed the trial state as a uniform product, the minimization of 
the variational ratio (Eq.(2$\cdot$3)) is equivalent to the minimization of the local 
energy for an arbitrary bond
\begin{equation}
\lambda( a, b, c ) = \frac{\langle \Psi | \left( -
S_{{\bf r}}^{\rm x} S_{{\bf r}'}^{\rm x} -
S_{{\bf r}}^{\rm y} S_{{\bf r}'}^{\rm y} + \alpha
S_{{\bf r}}^{\rm z} S_{{\bf r}'}^{\rm z}
\right) | \Psi \rangle}{\langle \Psi | \Psi \rangle} \, ,
\end{equation}
where ${\bf r}$ and ${\bf r}'$ are neighboring lattice points. In addition, the product 
structure of the trial state enables us to obtain 
the denominator
\begin{equation}
\langle \Psi | \Psi \rangle = \prod_{\bf r}^{~} \,\, \left[ W\left(
\begin{array}{cc} 
\sigma_{{\bf r}+{\hat {\bf j}}} & \sigma_{{\bf r}+{\hat {\bf i}}+{\hat {\bf j}}}\\ 
\sigma_{{\bf r}~~~}^{~} & \sigma_{{\bf r}+{\hat {\bf i}}~~~}
\end{array}
\right) \right]^2_{~} 
\end{equation}
as the partition function of the isotropic IRF model, which is specified by the local Boltzmann
weight $W^2_{~}$. Thus the norm of the trial state can be accurately calculated by 
use of DMRG applied to 2D classical systems.~\cite{Ni}
Similarly the numerator is also a partition function of an IRF model that has additional 
structure around the bond $\{ {\bf r} {\bf r}' \}$ in Eq.(2$\cdot$6), and can be calculated 
with sufficient numerical precision by DMRG as proposed by Hieida et al.~\cite{Hieida} 
We use CTMRG,~\cite{CTMRG1,CTMRG2} which is a variant of DMRG, 
for the calculations of the following result. 

\section{Calculated Result}

\begin{table}
\caption{Variational Energy and the value of optimal variational parameters.}
\label{table:1}
\begin{center}
\begin{tabular}{lcccccc} \hline \hline
~ & $4 \, \lambda_{\rm min}^{~}$ &
$S_{{\bf r}}^{\rm x} S_{{\bf r}'}^{\rm x}$ &
$S_{{\bf r}}^{\rm z} S_{{\bf r}'}^{\rm z}$ & a & b & c \\ \hline
Heisenberg & - 1.3089 & - 0.1245 & - 0.0782 & 0.8342 & 0.7483 & 0.5075 \\
XY & - 1.0848 & - 0.1356 & - 0.0383 & 0.9420 & 0.8796 & 0.6936 \\ \hline
\end{tabular}
\end{center}
\end{table}

The minimum of the variational energy $\lambda( a, b, c )$ in Eq.(2$\cdot$6) can be detected 
by way of the parameter sweep for $a$, $b$, and $c$. Table I shows the optimal parameter
sets that give the lowest variational energy $\lambda_{\rm min}^{~}$ for the 
isotropic Heisenberg model ($\alpha = 1$) and the XY model ($\alpha = 0$). 
The diagonal spin correlation $S_{{\bf r}}^{\rm z} S_{{\bf r}'}^{\rm z}$ and
the off-diagonal one $S_{{\bf r}}^{\rm x} S_{{\bf r}'}^{\rm x}$ between the
neighboring sites are also shown. In both cases the variational state is disordered, since there is
no phase transition from the parameter limit $a = b = c = 1$ to the 
shown parameter cases. Even at the Heisenberg point, the state is not
isotropic, as is observed from the difference 
between $S_{{\bf r}}^{\rm z} S_{{\bf r}'}^{\rm z}$ and
 $S_{{\bf r}}^{\rm x} S_{{\bf r}'}^{\rm x}$. 

\begin{table}
\caption{Energy expectation values.}
\label{table:2}
\begin{center}
\begin{tabular}{lcc} \hline \hline
~ & Heisenberg & XY \\ 
\hline
SM~\cite{Suz} & ~ & - 1.0743~ \\ 
TPVA & - 1.3089~ & - 1.0848~ \\
MC~\cite{MC} & - 1.33888 & - 1.09764 \\
\hline
\end{tabular}
\end{center}
\end{table}

Let us compare the calculated variational energy with two representative ground 
state energy estimations. Table II shows our result, a recent Monte Carlo (MC) result 
by Sandvik,~\cite{MC} and the variational energy calculated by Suzuki and 
Miyashita in 1978.~\cite{Suz} Our result for the Heisenberg (or the XY) model is 
2.3 \% (or 1.2 \%) higher than the MC result.

\section{Discussions}

As an application of DMRG for infinite size 2D quantum systems, we employ
TPVA for the square lattice $S = 1/2$ Heisenberg and XY models. The obtained 
variational state is disordered even for the Heiseberg point, the result which suggests 
the loss of the antiferromagnetic correlation.
The problem may be improved by increasing the variational parameter, as
Nishino {\it et al} have introduced auxiliary variables to the local weight when 
they applied TPVA to 3D classical models.~\cite{TPA1,VDM,TPA2}

\section*{Acknowledgments}

The authors thank to K.~Okunishi, Y.~Hieida, Y.~Akutsu, and S.~Miyashita for valuable 
discussions. The  
present research is partially supported by a Grant-in-Aid from Ministry of Education, 
Science and  Culture of Japan. A.~G is supported by Japan Socienty for the Promotion
of Science (P01192) and by VEGA grant No. 2/3118/23.


\begin{thebibliography}{99}
\bibitem{Wh1} S.R.~White, Phys. Rev. Lett. {\bf 69} (1992), 2863.
\bibitem{Wh2} S.R.~White, Phys. Rev. B {\bf 48} (1993), 10345.
\bibitem{rev} {\it Density-Matrix Renormalization --- A New Numerical
Method in Physics,} Lecture notes in Physics, eds. I.~Peschel, X.~Wang,
M.~Kaulke, and K.~Hallberg (Springer Verlag, 1999).
\bibitem{Shibata} N.~Shibata, J. Phys. A {\bf 36} (2003), 381.
\bibitem{Liang} S.~Liang and H.~Pang, Phys. Rev. B {\bf 49} (1994), 9214.
\bibitem{Xiang} T.~Xiang, J.Z.~Lou, and Z.B.~Su, Phys. Rev. B {\bf 64} (2001), 
104414.
\bibitem{Peshel1} M.C.~Chung and I.~Peschel, Phys. Rev. B {\bf 62} (2000), 4191.
\bibitem{Peshel2} M.C.~Chung and I.~Peschel, Phys. Rev. B {\bf 64} (2000), 064412.
\bibitem{CTTRG} T.~Nishino and K.~Okunishi, J. Phys. Soc. Jpn. {\bf 67}
(1998), 3066.
\bibitem{Ost} S.~\"Ostlund and S.~Rommer, Phys. Rev. Lett {\bf 75} (1995), 3537.
\bibitem{Rom}S.~Rommer and S.~\"Ostlund, Phys. Rev.  B {\bf 55} (1997), 2164. 
\bibitem{Sierra} J.~Dukelsky, M.A.~Mart\'{\i}n-Delgado, T.~Nishino and G.~Sierra, 
Europhys. Lett, {\bf 43} (1998), 457.
\bibitem{And} M.~Andersson, M.~Boman, and S.~\"Ostlund, Phys. Rev.  B {\bf 59}
(1999), 10493.
\bibitem{Sierra2} M.A.~Mart\'{\i}n-Delgado, M.~Roncaglia, and G.~Sierra, 
Phys. Rev. B {\bf 64} (2001), 075117.
\bibitem{KWA} H.A.~Kramers and G.H.~Wannier, Phys. Rev. {\bf 60} (1941), 263.
\bibitem{KWON} K.~Okunishi and T.~Nishino, Prog. Theor. Phys. {\bf 103} (2000), 541.
\bibitem{KWON2} T.~Nishino, K.~Okunishi, Y.~Hieida, N.~Maeshima and Y.~Akutsu, 
Nucl. Phys. B {\bf 575} (2000), 504.
\bibitem{Suz} M.~Suzuki and S.~Miyashita, Can. J. Phys. {\bf 56} (1978), 902.
\bibitem{Miy} S.~Miyashita, J. Phys. Soc. Jpn. {\bf 53} (1984), 44.
\bibitem{Hieida} Y.~Hieida, K.~Okunishi and Y.~Akutsu, New J. Phys. {\bf 1}
(1999), 7.
\bibitem{Zitt2} H.~Niggemann, A.~Kl\"umper  and J.~Zittartz, Z. Phys. 
{\bf B104} (1997), 103.
\bibitem{Ni} T.~Nishino,  J. Phys. Soc. Jpn. {\bf 64} (1995), 3598.
\bibitem{CTMRG1} T.~Nishino and K.~Okunishi, J. Phys. Soc. Jpn. {\bf 65}
(1996), 891.
\bibitem{CTMRG2} T.~Nishino and K.~Okunishi, J. Phys. Soc. Jpn. {\bf 66}
(1997), 3040.
\bibitem{MC} A.W.~Sandvik, Phys. Rev. B {\bf 60} (1999), 6588.
\bibitem{TPA1} T.~Nishino, Y.~Hieida, K.~Okunishi, N.~Maeshima and Y.~Akutsu,
Prog. Theor. Phys. {\bf 105} (2001), 409.
\bibitem{VDM} N.~Maeshima, Y.~Hieida, Y.~Akutsu, T.~Nishino, and K.~Okunishi, 
Phys. Rev. E {\bf 64} (2001), 016705.
\bibitem{TPA2} A.~Gendiar, N.~Maeshima, and T.~Nishino,
Prog. Theor. Phys. {\bf 110} (2003), 691.
\end{thebibliography}
\end{document}